\documentclass[12pt]{article}
\usepackage{epsfig}
\begin{document}
\begin{titlepage}

\begin{center}

\vspace{10cm}

  {\Large \bf  Ultra-light Glueballs in Quark-Gluon Plasma }
  \vspace{0.50cm}\\
  Nikolai Kochelev$^{a,b}$\footnote{Email address: kochelev@theor.jinr.ru}
\vskip 1ex

\it $^{a}$Institute of Modern Physics, Chinese Academy of Sciences, Lanzhou 730000, China
\vskip 1ex

\it $^{b}$ Bogoliubov Laboratory of Theoretical Physics, Joint
Institute for Nuclear Research, Dubna, Moscow region, 141980
Russia \vskip 1ex

\end{center}
\vskip 0.5cm \centerline{\bf Abstract}
We consider the dynamics of the scalar and pseudoscalar glueballs in the  Quark-Gluon Plasma (QGP).
By using   the instanton model for the QCD vacuum we give the arguments
 that the nonperturbative gluon-gluon interaction   is qualitatively  different
 in the confinement and deconfinement phases. Based on this observation
 it is shown that above  $T_c$ the values of the scalar and pseudoscalar
 glueball masses might be very small.
The estimation of the temperature of scale invariance restoration, at which the scalar glueball
becomes massless, is given. We also discuss the
  Bose-Einstein condensation of the glueballs and the superfluidity of the glueball matter  in QGP.
\end{titlepage}

\setcounter{footnote}{0}
\section{Introduction}

In spite of the tremendous experimental and theoretical  efforts in the investigation of   the
properties of quark-gluon plasma (QGP) at large temperatures and densities, it is not  clear so far
 what is the
fundamental QCD mechanism leading to the unusual behaviour of the matter produced in heavy ion
collisions at high energies at the RHIC and LHC.
Indeed, there is strong evidence that even above $T_c$ non-perturbative QCD effects are very important
and, in particular, they are responsible for the phenomenon  of the so-called
strongly-interacting Quark-Gluon Plasma (sQGP) \cite{shuryakQGP} observed at the RHIC and LHC.
One of the important outputs of the Lattice QCD calculation at finite T is  the fact that
the values of the deconfinement and chiral restoration temperatures are approximately equal.
That means that above $T_c$ one cannot expect the pion to be the  Goldstone boson with
zero mass in the chiral
limit. Instead,  a rather massive ($ M_{\pi}\approx 2M_q(T_c)\approx 6T_c \sim 1$ GeV)
pion  state  appears  above $T_c\approx 150$ MeV. Therefore,
this lowest mass quark-antiquark state can not give a significant contribution to the Equation
of State (EoS) of the QGP.

One of the fundamental issues  of QCD as the
theory of strong interactions  is the understanding of the
 role of gluonic degrees of freedom in the confinement
and deconfinement regimes.
Thus, it is known that at low temperatures gluons play an important role not only in the
dynamics of usual hadrons. In particular,
 they can form bound states
called glueballs (see review \cite{Mathieu:2008me}).
However, one cannot expect a significant contribution of glueballs at $T<T_c$ to
EoS of the hadron gas  due to
their large masses $M_G\gg T$.
On the other hand, it was found by lattice calculation \cite{lattice1,Miller:2006hr} that unlike the quark condensate,  the gluon condensate
does not vanish at $T>T_c$. We should stress that this condensate plays a
 fundamental role in the  formation of the
bound glueball states. This role is similar to the role of the quark condensate in the appearance of the
massive  mesons and baryons built of the light $u$, $d$, and $s$  quarks.
Therefore, a non-zero value  of the gluon condensate above $T_c$  is
 a strong signal of the existence of the  glueballs in the deconfinement phase.

However, it is evident that the properties of the glueballs, in particular, their masses and sizes
should  change in the QGP due to the temperature dependence  of the  gluon condensate \cite{top,Miller:2006hr}
and the change of the topological structure of the QCD vacuum at $T>T_c$ \cite{top,top2}.
The attempts to estimate the value of the glueball masses at finite $T$ were done
in the different models in the papers
 \cite{Vento:2006wh,Kochelev:2006sx,Lacroix:2012pt,Buisseret:2009eb,Agasian:1993fn}
and in the lattice
\cite{Meng:2009hh,Caselle:2013qpa}.

In this Letter, we consider the lowest scalar and
pseudoscalar  glueball states at finite $T$ in the QGP environment within the
effective  model
based on the instanton picture for the QCD vacuum.
It is shown that their masses strongly decrease above $T_c$ and
become very small at the temperature of the scale invariance
restoration  $T_{scale}\approx 1$ GeV.
The possibility
of the Bose-Einstein condensation and superfluidity  of the  glueball matter is under discussion.

\section{Nonperturbative quark-quark, quark-gluon and gluon-gluon interactions below and
above $T_c$ induced by instantons}

It is expected that the instantons, a strong vacuum fluctuation of gluon
fields, play a very important role in the dynamics of the glueballs below $T_c$
(see review \cite{shuryak}).
The effective interaction induced by the instanton between quarks at $T=0$ is well known.
This is a famous t'Hooft interaction which
for $N_f=3$~(Fig.~1a) and $N_c=3 $   is given by
the formula \cite{Shifman:1979uw}\footnote{ In this section, all Lagrangians are written in the Euclidian
space-time.}:
\begin{eqnarray}
{\cal L}_{eff}^{(3)}&=&\int d\rho\
 n(\rho)\bigg\{\prod_{i=u,d,s}
\bigg(m_i^{cur} \rho-\frac{4\pi}{3}\rho^3\bar q_{iR}q_{iL}\bigg)
\nonumber\\
&&+\frac{3}{32}\bigg(\frac{4}{3}\pi^2\rho^3\bigg)^2
\bigg[\bigg(j_u^aj_d^a-\frac{3}{4}j_{u\mu\nu}^a
j_{d\mu\nu}^a\bigg)\bigg(m_s^{cur}\rho-\frac{4}{3}\pi^2\rho^3\bar
q_{SR}q_{sL}\bigg)\nonumber\\
&&+\frac{9}{40}\bigg(\frac{4}{3}\pi^2\rho^3\bigg)^2d^{abc}
j_{u\mu\nu}^aj_{d\mu\nu}^b
j_s^c+ {\rm perm.}\bigg]+\frac{9}{320}\bigg(\frac{4}{3}\pi^2\rho^3\bigg)^3
d^{abc}j_u^aj_d^b
j_s^c\nonumber\\
&&+\frac{if^{abc}}{256}\bigg(\frac{4}{3}\pi^2\rho^3\bigg)^3
j_{u\mu\nu}^aj_{d\nu\lambda}^b\relax
j_{s\lambda\mu}^c+(R\longleftrightarrow L) \bigg\} ,
\label{thooft3}
\end{eqnarray}
where, $m_i^{cur}$ is the quark current mass,
$q_{R,L}={(1\pm\gamma_5)q(x)/2}, \ j_i^a=\bar
q_{iR}\lambda^aq_{iL},{\ } j_{i\mu\nu}^a=\bar
q_{iR}\sigma_{\mu\nu}\lambda^aq_{iL}$,
 $\rho$ is the instanton size and  $n(\rho)$ is the density of
instantons.
For  $N_f=2$ the t'Hooft interaction for zero current quark mass is much simpler
\begin{eqnarray}
{\cal L}_{eff}^{(N_f=2)}&=&\int d\rho n(\rho)(\frac{3}{4}\pi^2\rho^3)^2\bar q_{iR}q_{iL} \bar
q_{jR}q_{jL}
\bigg[1+\frac{3}{32}\lambda_u^a\lambda_d^a\nonumber\\
&&+\frac{9}{32}\vec{\sigma_u}\cdot\vec{\sigma_d}\lambda_u^a\lambda_d^a\bigg]
+(R\longleftrightarrow L). \label{thooft2}
\end{eqnarray}
This Lagrangian can be obtained from Eq.(\ref{thooft3}) by connecting the strange quark legs
through the quark condensate.
The two Lagrangians  Eq.(\ref{thooft3}) and Eq.(\ref{thooft2}) are considered as the bases
for different versions of the Nambu-Jona-Lasinio (NJL)   models \cite{Diakonov:2002fq}.

The effective quark-gluon
 interaction induced by  the instantons
   is (see, for example \cite{Zetocha:2002as,ABC})
\begin{eqnarray}
{\cal L}_{eff}&=&\int dUd\rho n(\rho)\prod_q
-2\pi^2\rho^3
  \bar q_R(1+\frac{i}{4}U_{ab} \tau^a
\bar\eta_{b\mu\nu}\sigma_{\mu\nu})q_L
\nonumber\\
&\times&
e^{-\frac{2\pi^2}{g_s}\rho^2U_{cd}\bar\eta_{d\alpha\beta}G^c_{\alpha\beta}}+(R\leftrightarrow
L), \label{gluons}
\end{eqnarray}
where
  $U$ is
  the orientation matrix of the instanton in the $SU(3)_c$ color space.
From this Lagrangian one can obtain the
interaction between quarks and gluons which contributes to the
mixing between scalar and pseudoscalar glueballs, quarkoniums and tetraquark states
\begin{equation}
{\cal L}_{ggq\bar q}=\int d\rho \frac{\pi^3\rho^4 n(\rho)}{8\alpha_s(\rho)}
\bigg[(G^2-G\widetilde{G}){\cal L}_{f,I}+(G^2+G\widetilde{G}){\cal L}_{f,A}\bigg],
\label{lagq}
\end{equation}
where
\begin{equation}
G^2\equiv G_{\mu\nu}^a G_{\mu\nu}^a,\ \ \
G\widetilde{G}\equiv G_{\mu\nu}^a \widetilde{G}_{\mu\nu}^a,
 \label{gqq}
\end{equation}
where $\widetilde
G_{\mu\nu}^a(x)=1/2\epsilon_{\mu\nu\alpha\beta}G_{\alpha\beta}^a(x)$, and
 ${\cal L}_{f,I}$ and  ${\cal L}_{f,A}$ are the t'Hooft  interaction
 induced by the instanton and antiinstanton, respectively.

\begin{figure}[htb]
\vspace*{-0.0cm} \centering
\centerline{\epsfig{file=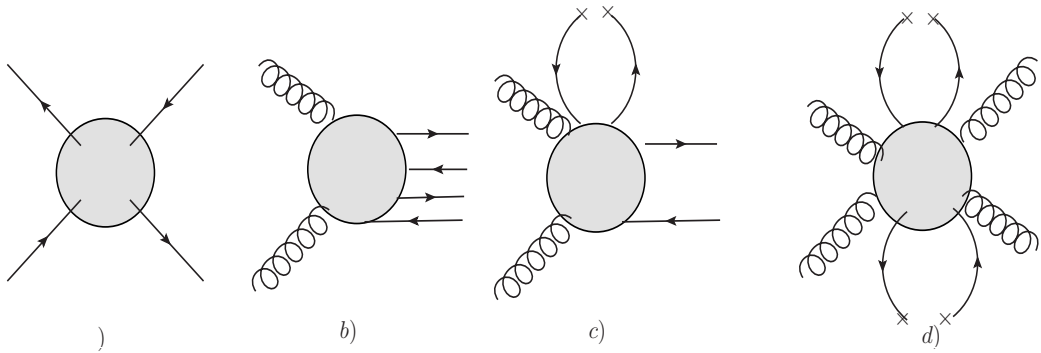,width=14cm,height=5cm, angle=0}}\
\caption{ The quark-quark a), quark-gluon b), c) and gluon-gluon interaction d)
induced by the instanton for the $N_f=2$ case.
The symbol $\times$ means connection through the quark  condensate or  by the currect quark mass.}
\end{figure}

 The  interaction between gluons in the scalar and pseudoscalar channels induced by the instanton
  is
\begin{equation}
{\cal L}_{gg}=\frac{\pi^6}{80}\int d\rho \frac{n(\rho)\rho^8}{\alpha_s^2(\rho)}\bigg[
 G_{\mu\nu}^a G_{\mu\nu}^aG_{\sigma\tau}^b G_{\sigma\tau}^b+
 G_{\mu\nu}^a \widetilde{G}_{\mu\nu}^aG_{\sigma\tau}^b \widetilde{G}_{\sigma\tau}^b   \bigg].
  \label{gg}
\end{equation}

Due to factor $2\pi^2/g_s$ in Eq.(\ref{gluons}) each two gluon legs produce
a large enhancement factor $\pi^6/\alpha_s^2$ in the gluon-gluon  interaction induced
by instantons. So one can expect that this type of the interaction is much stronger in
 comparison
with the quark-quark case.

We would like to emphasize  that the instanton induced interaction
is very sensitive to the parity of the glueball and quarkonium states
and, in particular, it leads  to the mass splitting between
scalar and pseudoscalar glueballs \cite{forkel,kochmin}.
Indeed, the single
instanton contribution to the difference of two correlators of
glueball currents with opposite parities  is given by
\begin{eqnarray}
\Delta\Pi(Q^2)_G&=&i\int
d^4xe^{iqx}(<0|TO_S(x)O_S(0)|0>-<0|T|O_P(x)O_P(0)|0>) \nonumber\\
&=&2^6\pi^2 \int d\rho n(\rho)(\rho Q)^4{K_2}^2(\rho
Q), \label{direct}
\end{eqnarray}
where the glueball currents for the scalar and pseudoscalar states are
the following:
\begin{eqnarray}
O_S(x)&=&\alpha_sG_{\mu\nu}^a(x)G_{\mu\nu}^a(x),\label{pseudo}\\
O_P(x)&=&\alpha_sG_{\mu\nu}^a(x)\widetilde
G_{\mu\nu}^a(x).\label{scalar}
\end{eqnarray}
\begin{figure}[htb]
\vspace*{-0.0cm} \centering
\centerline{\epsfig{file=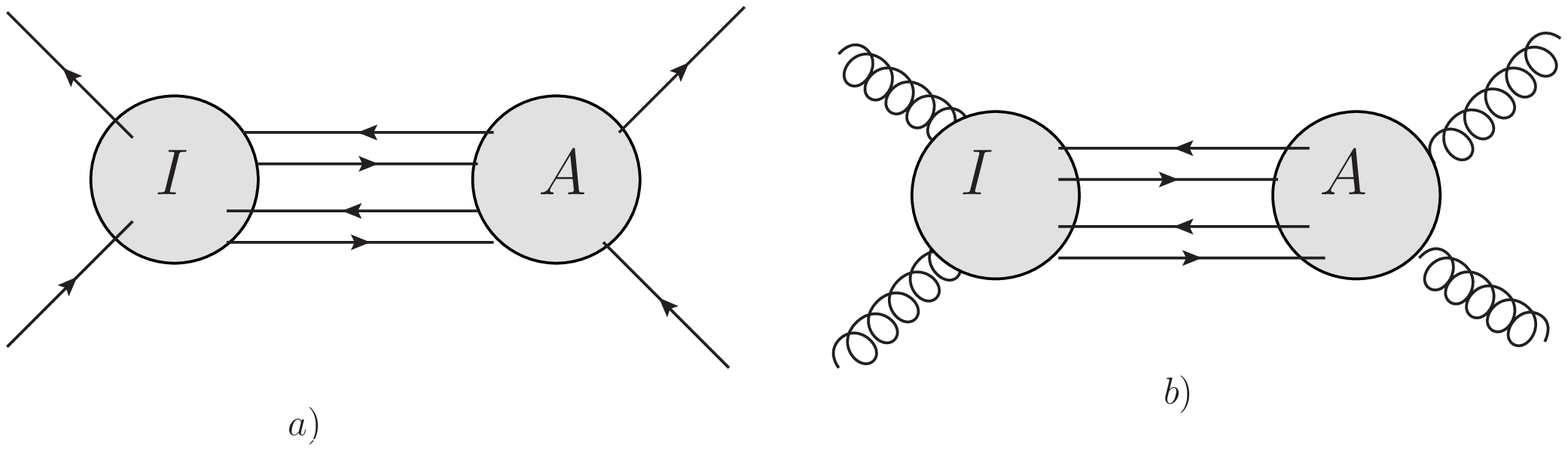,width=14cm,height=5cm, angle=0}}\
\caption{ The quark-quark a) and gluon-gluon interaction b)
induced by instanton-antiinstanton molecule.}
\end{figure}

At $T<T_c$ the main contribution to the quark-quark and gluon-gluon   interaction
 comes from  disordered instantons, Fig.1,  in the
 so-called "random" phase of the instanton liquid. The phase provides
 spontaneous chiral symmetry breaking in QCD (see reviews \cite{shuryak,Diakonov:2002fq}).
In this case, the  chiral symmetric contribution,  which
is related to the  correlated instanton-antiinstanton molecules, Fig.2, is expected to be small
due to the small packing fraction  of instantons in the QCD vacuum \cite{shuryak}.
Above $T_c$, where the
chiral  and $U_A(1)$ symmetries are restored, the situation is
opposite. Indeed,  in this case the contribution from the chirality violated
interaction, Fig.1, is  proportional to the product of the current quark masses and should be small.
In the chiral symmetric phase,
  the leading  contribution comes from the strongly correlated
 instanton-antiinstanton molecules, Fig.2, as  supported by the   calculations
in \cite{Ilgenfritz}.
\begin{figure}[htb]
\vspace*{-0.0cm} \centering
\centerline{\epsfig{file=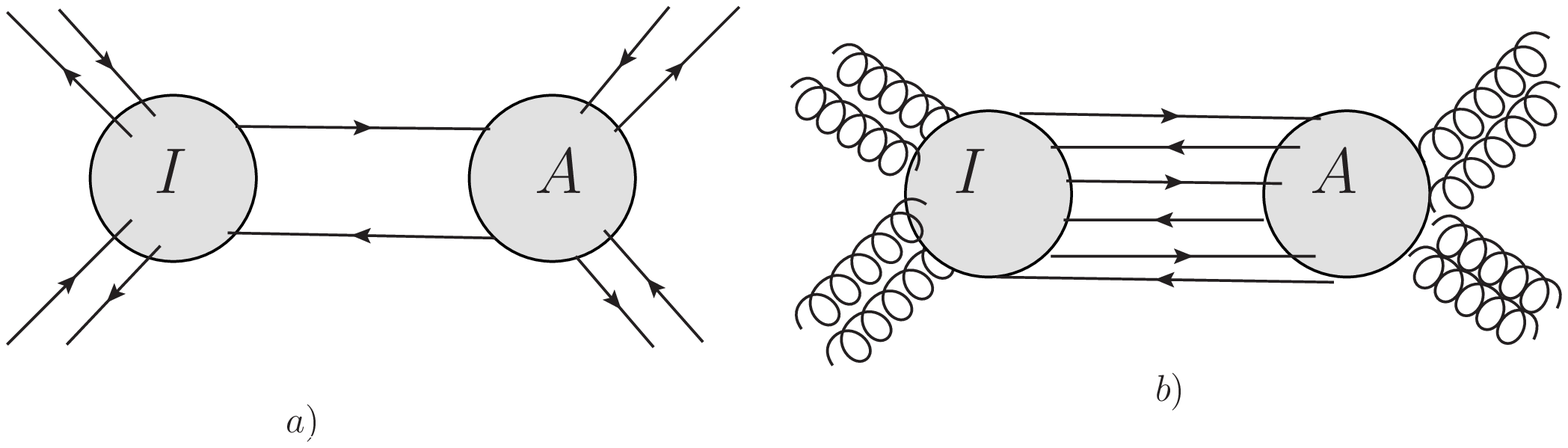,width=14cm,height=5cm, angle=0}}\
\caption{ The eight-quark a) and eight-gluon interactions  b)
induced by instanton-antiinstanton molecule for the $N_f=3$ case.}
\end{figure}
The contribution from the eight-quark interaction presented in Fig.3a does
not give a significant contribution to the hadron spectroscopy but it is important for
the stability of the vacuum of the NJL model with the instanton induced interaction
\cite{Osipov:2005tq,Osipov:2006ns}.

The effects of quark-quark interaction induced by instanton-antiinstanton molecules
near $T_c$ were studied in \cite{rho,Brown:2004jy}. It was shown that they might give
a significant contribution to the binding energy of  the quark-antiquark
states in the deconfinement
phase.
The main effect, in comparison with the zero temperature case, comes  from the
strong polarization of instanton-antiinstanton molecules in the time direction at the high
temperature.

\section{Lowest mass scalar glueball above~$T_c$ }

At $T=0$ there are several  scalar meson states   $f_0(600)$, $f_0(980)$,
   $f_0(1500)$ and $f_0(1710)$   with a  possible admixture of  gluons
 \cite{Mathieu:2008me}\footnote{We do not include in this list the $f_0(1370)$ state due to
 its controversial  situation in the experiment (see the recent discussion in  \cite{Ochs:2013vxa}).}.
Unfortunately, even the  lattice calculation, which is  based on the first principles of QCD,
  cannot give an accurate prediction for the  scalar glueball masses with unquenched quarks.
The main problem here  is in the large contribution coming  from the so-called disconnected diagrams
\cite{Wakayama:2014gpa}.
 In particular, they might also be  responsible  for the  strong
 quarkonium-glueball mixing. Within the instanton model such mixing was shown to be large
 in the different approaches in  \cite{Kochelev:2005tu},\cite{Harnett:2008cw}.
Therefore, it is quite difficult, even  impossible, to find a pure glueball state at $T=0$.
Several arguments based on the analysis
of sigma-meson contribution to different reactions
   were  given to consider  the lowest mass scalar
  sigma-meson $f_0(600)$ state as the state with
 a large mixture of the
glueball \cite{Kaminski:2009qg,Mennessier:2010ij,Ochs:2013gi}.
In the line of these studies, we will also assume that at  $T=0$ the lightest glueball state is
the sigma meson state with the mass $m_\sigma \approx 450 $ MeV.
It is also well known that the masses of the hadron states  are changing only a little
in the interval $0<T<T_c$ due to a small change of the values of quark and gluon
condensates in this region.
Therefore, the mass of sigma should be
around $450$ MeV  at $T\approx T_c$.
It has been  discussed above that  in the  chiral symmetric phase above $T_c$ the main
contribution to the gluon-gluon interaction is  related to the formation of  the instanton-antiinstanton
molecules. The example of these interactions is presented in Fig.2b and Fig.3b.
Similarly to the case  of the NJL model with the eight-quark interaction based on  the instanton
induced multiquark interaction \cite{Osipov:2005tq,Osipov:2006ns},
the effective Lagrangian in the gluon sector has the form
\begin{eqnarray}
{{\cal L}(\Phi)}=\frac{1}{2}(\partial_\mu\Phi)^2-V(\Phi), \label{pot}
\end{eqnarray}
where
\begin{eqnarray}
V_0(\Phi)=-\frac{\mu^2}{2}\Phi^2+\frac{\lambda}{4}\Phi^4, \label{lambda}
\end{eqnarray}
and $\Phi$ is now the scalar glueball field.

 Due to the scale anomaly in QCD, the trace energy momentum tensor is  not zero
but it is proportional to the value of the gluon condensate
\begin{eqnarray}
{T^\mu}_\mu=-\frac{9}{8}<\frac{\alpha_s}{\pi}G^2>. \label{scale}
\end{eqnarray}

By using the  parton-hadron duality principle
we obtain the relation between the gluon condensate in QCD and the minimum of the
potential, Eq.\ref{lambda}, in the glueball sector
\begin{eqnarray}
{T^\mu}_\mu=4V(\Phi_0). \nonumber
\end{eqnarray}
To estimate the coupling $\lambda$,
  we use the relation
\begin{eqnarray}
\lambda=\frac{2\pi m_0^4}{9<\alpha_s G^2>}, \nonumber
\end{eqnarray}
and the value of the gluon condensate \cite{Narison:2011rn}
\begin{eqnarray}
<\alpha_s G^2>\approx 0.07 GeV^4. \nonumber
\end{eqnarray}
For the  mass of the glueball at $T\approx T_c$ $m_0\approx 450$ MeV we have
$\lambda=0.41$. Now we are in a position to calculate the temperature dependence
of the scalar glueball mass above $T_c$.
At the finite temperature, the effective potential in the one loop approximation
for the $\lambda\Phi^4$ theory is given (see for example, \cite{kirzhnits})
\begin{equation}
V(\Phi(T))=V_0(\Phi(T))+\frac{3\lambda\Phi(T)^2}{4\pi^2}\int \frac{k^2dk}
{\sqrt{k^2+m_0^2}(exp(\sqrt{k^2+m_0^2}/T)-1)}\label{poten}
\end{equation}
By using the condition of the minimum of $V(\Phi(T))$

\begin{equation}
\frac{\delta V(\Phi(T))}{\delta \Phi(T)}=0,
\label{potenprime}
\end{equation}

and the  definition for the mass of the glueball
\begin{equation}
\frac{\delta^2 V(\Phi(T))}{\delta\Phi(T)^2}=m^2_\Phi(T)\nonumber,
\end{equation}
we obtain
\begin{equation}
m^2_\Phi(T)=m_0^2-\frac{3\lambda}{\pi^2}\int \frac{k^2dk}
{\sqrt{k^2+m_0^2}(exp(\sqrt{k^2+m_0^2}/T)-1)}\label{mass},
\end{equation}
and
\begin{equation}
m^2_\Phi(T)=2\lambda\Phi^2_{min}(T).
\label{phi}
\end{equation}

\begin{figure}[htb]
\vspace*{-0.0cm} \centering
\centerline{\epsfig{file=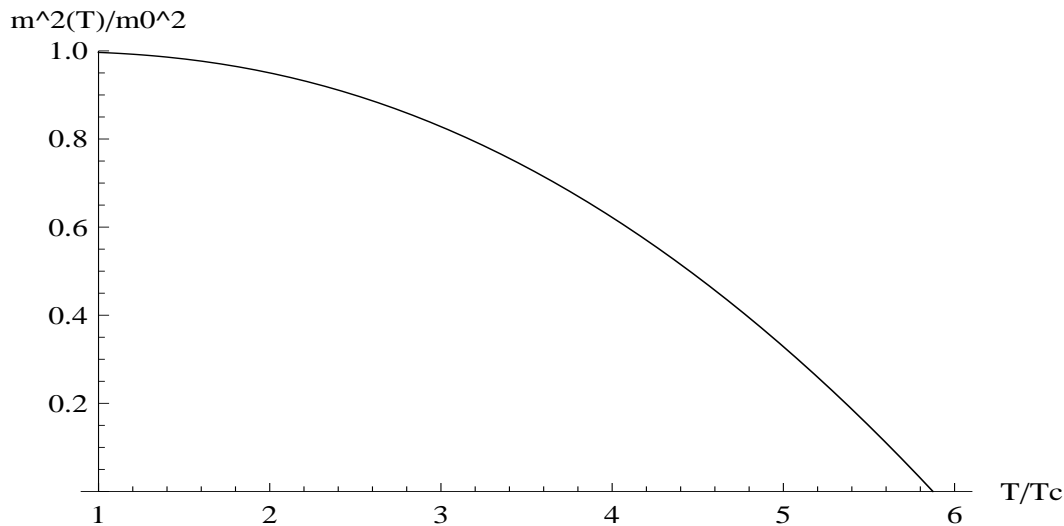,width=14cm,height=7cm, angle=0}}\
\caption{ The ratio of $m^2_\Phi(T))/m_0^2$ as a function of $T/T_c$.}
\label{ratio}
\end{figure}

For the trace anomaly we obtain
\begin{eqnarray}
{T^\mu}_\mu=4V(\Phi_{min}(T))=-\lambda\Phi^4_{min}(T)=-\frac{m^4_\Phi(T)}{4\lambda}.\nonumber
\end{eqnarray}
The temperature dependence of the gluon condensate in this model is given by the  formula
\begin{equation}
<\frac{\alpha_s}{\pi}G^2(T)>=\frac{2m^4_\Phi(T)}{9\lambda}
\label{condensate}
\end{equation}
In Fig.\ref{ratio}, the temperature dependence of the mass of the scalar glueball
is presented.  So one can see that at $T_{scale}\approx 6T_c\approx 0.9 $ GeV the mass of the scalar glueball
 vanishes and the scale invariance is restored.
This can be treated as the appearance of the massless dilaton field in the limit ${T^\mu}_\mu\rightarrow 0$
(see the recent discussion of the dilaton in \cite{Crewther:2014kha}).

We should emphasize  that the justification of the use of the point-like effective  interaction in
 Eq.(\ref{lambda})  can be related to a
very small size of the scalar glueball in the QGP. Even at $T=0$, due to the strong attraction
between gluons
induced by the instantons, the size of the scalar glueball is very small,
$R_g\approx 2/3\rho_c\approx 0.2$ fm, where $\rho_c\approx 0.3$ fm is the average instanton size in the QCD
vacuum \cite{Schafer:1994fd}. This small size was also confirmed in the lattice calculation
\cite{deForcrand:1991kc}, \cite{Liang:2014jta}. Above $T_c$ a similar phenomenon happens as well.
However, in this case the
instanton-antiinstanton molecules  produce a very strong attraction in the  scalar glueball channel.
Therefore, we expect that the size of the scalar glueball should
be smaller than the  size of instantons in the QGP, which is cutting at
\begin{equation}
R_G\ll\bar\rho^2(T)\approx \frac{1}{3\pi^2T^2},
\label{size}
\end{equation}
at finite T and $N_c=N_f=3$ \cite{pisarski}.
We would like to emphasize that this size is smaller than the perturbative
 Debye screening length in the QGP
\begin{equation}
\lambda_D^2(T)=1/M_D^2(T)>\frac{1}{3\pi T^2},
\label{size2}
\end{equation}
where    $\alpha_s(\rho_c)\approx 0.5$  \cite{Diakonov:2002fq} for $T=0$ and
$M_D^2(T)=g_s^2(N_c/3+N_f/6)T^2$  \cite{shuryak1978} was used.
Therefore, we come to the
important conclusion that the  Debye screening cannot destroy
the binding of the scalar glueball in QGP.

\section{Pseudoscalar glueball  above~$T_c$ }

There are also several candidates for the pseudoscalar glueball below $2$ GeV  at $T=0$:
$\eta(1405), \eta(1475), \eta(1760)$ and $X(1835)$. In the lattice quenched  calculation the lowest
mass pseudoscalar glueball was predicted to have the mass $M_{0^{-+}}\approx 2.6$ GeV
\cite{lattice3,Chen:2005mg}\footnote{ The difference between the  observed and lattice masses might
be related to the strong mixing between the glueball and quarkonium states \cite{Kochelev:2005tu}.}.
The large difference between scalar and pseudoscalar glueball masses
$m_{0^{-+}}(T=0)-m_{0^{++}}(T=0)\approx 1 $ GeV at low temperature can be explained by the large
single instanton contribution to these channels
(see discussion in  \cite{shuryak}).
Indeed, it gives a strong attraction in the scalar channel and a strong repulsion in the pseudoscalar state.
After restoration of the chiral symmetry the masses of scalar and pseudoscalar glueballs
should be equal in the limit of the zero  light quark masses, $m_u=m_d=m_s=0$. In this case,
instanton-antiinstanton molecules give the same strength of attraction in the both states.
Therefore, the leading contribution to the splitting  between the masses of two states at $T>T_c$ is determined by
the density of the single instantons,  which is proportional
to the product of the current masses of the light quarks.
We should mention that even at large temperature the single instanton contribution is repulsive
in the pseudoscalar glueball case, and its collapse to the massless state is not allowed.
This is a very important difference between the properties of pseudoscalar and scalar glueballs
at the temperature of the scale invariance restoration.

 We can estimate the mass of the pseudoscalar glueball at $T=T_{scale}$
by using the instanton model. The difference between the scalar and pseudoscalar glueball masses above $T_c$
is determined by the instanton density $n(\rho(T))$
\begin{equation}
n(\rho(T))\sim m_u m_d m_s\bigg(\rho(T)\bigg )^{b_0-2},
\label{den}
\end{equation}
where $b_0=11N_c/3-2N_f/3$ and $m_i$ are the current masses of the quarks.
Finally, we have
\begin{equation}
m_{0^{-+}}(T_{scale})\approx \frac{m_u m_d m_s}{m^*_u m_d^* m^*_s}
\bigg (\frac{1}{\sqrt{3}\pi T_{scale}\rho_c}\bigg)^7(m_{0^{-+}}(T=0)-m_{0^{++}}(T=0)),
\nonumber
\end{equation}
where $ m^*_i=m_i+m^*$ is the so-called effective mass of the quark in the instanton vacuum related to the
quark condensate.
The use of $m_u=m_d\approx 4.5$ MeV, $m_s\approx 100$  MeV \cite{Narison:2014wqa},
  $m^*=170$ MeV \cite{shuryak}, $\rho_c=1/600$ MeV $^{-1}$
   and $m_{0^{-+}}(T=0)-m_{0^{++}}(T=0)\approx 1 $ GeV
gives the estimation

\begin{equation}
m_{0^{-+}}(T_{scale}) \approx 0.1 \ \  eV.
\nonumber
\end{equation}
Therefore, the mass of the pseudoscalar glueball at $T=T_{scale}$ is finite but very small.

\section{ Bose-Einstein condensation of the  light glueballs  and superfluidity of glueball matter in QGP
}
\begin{figure}[htb]
\vspace*{-0.0cm} \centering
\centerline{\epsfig{file=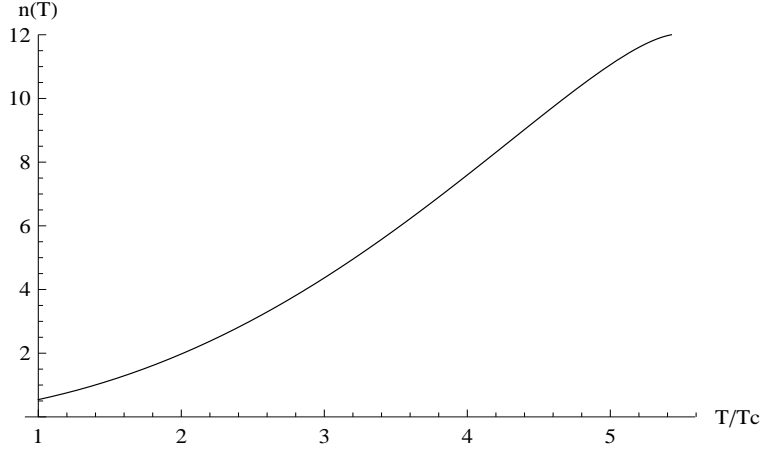,width=10
cm,height=6cm, angle=0}}\
\caption{ The T-dependency of the BEC density of glueballs in the $1/fm^3$ units.}
\label{BECfigure}
\end{figure}

The Bose-Einstein condensation (BEC) of the identical bosons
 is the well-known phenomenon in both theoretical and experimental physics.
Recently, it was suggested that the over-occupied  initial state of gluons created during
relativistic heavy ion collisions might lead to the BEC of gluons \cite{Blaizot:2011xf,
Blaizot:2013lga,Scardina:2014gxa,Xu:2014ega}. However,
gluons have the color charge and spin and, additionaly, the number of gluons is not conserved. All
of these features, lead to a very complicated study of the possible
formation of the gluon  BEC in the QGP.
On the other hand, the light glueballs with zero spin and color can easily form BEC at rather high temperature
because  for the  given boson number density $n$  the BEC temperature depends on the mass as \cite{kapusta}
\begin{equation}
T_{BEC}\approx 3.31 \frac{n^{2/3}}{m},
\label{nonrel}
\end{equation}
for the non-relativistic case.
For the arbitrary boson mass the critical BEC density is related to the temperature by the equation (see, for example,
 \cite{Grether:2007ur} and references therein)
\begin{equation}
n_{BEC}= \int\frac{d^3k}{(2\pi)^3}\frac{1}{exp((\sqrt{k^2+m^2}-m)/T_{BEC})-1}.
\label{BECREL}
\end{equation}
In Fig.\ref{BECfigure}, we show the critical glueball density of the scalar
and pseudoscalar glueballs in the QGP  by using the T-dependent glueball mass from
Eq.(\ref{mass}) \footnote{ We neglect a tiny difference between the values of
the scalar and pseudoscalar glueballs in the QGP.}.
This density can be compared with the  density of QGP  at the time $\tau$ produced at the RHIC, which was estimated by using the
multiplicity of the meson production in the Au-Au central collision \cite{hydro}
\begin{equation}
n_{RHIC}\approx \frac{13}{\tau} (fm^{-3}),
\label{RHICden}
\end{equation}
where $\tau$ is in $fm$. For LHC Pb-Pb central collisions at $\sqrt{s_{NN}}=2.76 $ GeV
the multiplicity is approximately twice larger. Therefore, the density is
\begin{equation}
n_{LHC}\approx \frac{26}{\tau} (fm^{-3}).
\label{LHCden}
\end{equation}
For QGP in the thermal equilibrium  we can estimate the contribution of the glueballs to
the total density. In this case, we have approximately 16 gluon and 24 quark degrees of freedom ($N_f=2$)
and 2 for glueballs. For the gluons and quarks with the mass $M_{qg}\approx 3T$ \cite{Petreczky:2001yp},
we have
for  scalar (pseudoscalar)  glueball contribution  to the total density at the  RHIC
\begin{equation}
n_{RHIC}^{glueball} \approx \frac{3}{\tau} (fm^{-3}),
\label{gludensRHIC}
\end{equation}
for the initial temperature $T_0^{RHIC} \approx 300 $ MeV.

For the LHC energy  ( $T_0^{LHC}\approx 400$ MeV)  the estimation is
\begin{equation}
n_{LHC}^{glueball} \approx \frac{5}{\tau} (fm^{-3}).
\label{gludensLHC}
\end{equation}

\begin{figure}[htb]
\vspace*{-0.0cm} \centering
\centerline{\epsfig{file=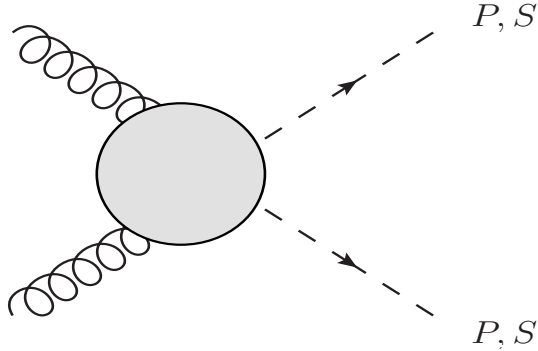,width=8
cm,height=5cm, angle=0}}\
\caption{ The production of the scalar $S$ and pseudoscalar $P$ glueballs in the  nonequilibrium QGP.}
\label{production}
\end{figure}
By using the Bjorken model for the QGP expansion \cite{Bjorken:1982qr}
\begin{equation}
\tau T^3=\tau_0 T_0^3,
\label{temp}
\end{equation}
with the thermalization time $\tau_0\approx 1$ fm,
we can estimate BEC temperature for the RHIC and LHC
\begin{equation}
T_{BEC}^{RHIC}\approx 200 MeV,  \ \ \ T_{BEC}^{LHC}\approx 270 MeV.
\label{BECthem}
\end{equation}
Therefore, the formation of the glueball BEC is possible at both RHIC and LHC heavy ion collisions.
We should also mention that there is a large gap between the mass of the light glueball and the mass of the
excited state in the system of two gluons in the QGP, which is $m_{gg}\approx 6T> 1 GeV$.
It is well know that such a gap should lead to the phenomenon of the superfluidity of the BEC matter.
So we arrive at the conclusion that the QGP at the RHIC and LHC might be considered as the mixture of
three matters. One of them  is the usual "normal" QGP matter consisting of quarks and gluons and
other ones are two superfluid matters of very light scalar and pseudoscalar glueballs.

There is also an additional mechanism which can lead to the abundance of the light glueballs
production in ultra-relativistic heavy ion collisions. Indeed, at very high temperatures at the
initial stage of the production of the quark-gluon matter, the pair production of glueballs is possible
by the fusion of two gluons, Fig.6.  The effective interaction responsible for such production
is
\begin{equation}
{\cal L}=\lambda_2(T)\alpha_sG_{\mu\nu}^a G_{\mu\nu}^a(S^2+P^2),
\label{inter}
\end{equation}
where $S(P)$ are the scalar (pseudoscalar) fields.
In the mean field approximation we have
\begin{equation}
\lambda_2(T) \sim \frac{m_{S,P}(T)^2}{\alpha_sG_{\mu\nu}^a G_{\mu\nu}^a(T)}\sim \frac{1}{m^2_{S,P}(T)}.
\label{la2}
\end{equation}
Therefore, one might expect a strong enhancement of the ultra-light glueball
 production in the phase
where  the quark-gluon plasma is far away from the  equilibrium.

\section{Conclusion}

In summary, we considered the properties of scalar and pseudoscalar
glueballs in the Quark-Gluon Plasma created in relativistic heavy-ion collisions.
Based on the instanton model for the QCD vacuum  we gave the arguments in favor
of the existence of very light scalar and pseudoscalar glueball states above the
temperature of the deconfinement transition.
The estimation of the temperature of the scale invariance restoration, at which the scalar glueball
becomes massless, was given. We also discussed the mechanism of the
  Bose-Einstein condensation  and the superfluidity of the scalar and pseudoscalar
  glueball matter  in the QGP. We also showed the possibility of the abundance glueball
  production at the initial stage of the QGP formation. The influence of this phenomenon on the
  fast thermalization of the QGP is the subject of our future investigation.

\section{Acknowledgments}

I would like to thank A.E. Dorokhov, A.Di Giacomo, S.B. Gerasimov, D.G. Pak, M.-E.Ilgenfritz, M. Oka,
Hee-Jung Lee, Yongseok Oh and Pengming Zhang
   for useful
discussions. I am  very grateful to Vicente Vento for numerous discussions of the properties
of the glueballs in the QGP.
I acknowledge cordial hospitality by the Institute of Modern Physics, CAS in Lanzhou,
where this work was completed.
This work was partially supported by the
National Natural Science Foundation of China (Grant No. 11035006 and 11175215),
and by the Chinese
Academy of Sciences visiting professorship for senior international scientists (Grant No. 2013T2J0011).
The work was initiated through the series of APCTP-BLTP JINR Joint Workshops.

\end{document}